\documentclass[Physsubmission, Phys]{SciPost}

\hypersetup{
    colorlinks,
    linkcolor={red!50!black},
    citecolor={blue!50!black},
    urlcolor={blue!80!black}
}

\usepackage[bitstream-charter]{mathdesign}
\DeclareSymbolFont{usualmathcal}{OMS}{cmsy}{m}{n}
\DeclareSymbolFontAlphabet{\mathcal}{usualmathcal}

\def\powhegbox{\textsc{Powheg-Box}}

\def\mg5{\textsc{MG5\_aMC@NLO}}

\def\sherpa{\textsc{Sherpa}}
\def\nlox{\textsc{Nlox}}
\def\helac{\textsc{Helac-Nlo}}
\def\GeV{\textrm{GeV}}
\def\dipoles{\textsc{Helac-Dipoles}}
\def\1loop{\textsc{Helac-1Loop}}
\def\lo{\textrm{LO}}
\def\nlo{\textrm{NLO}}

\begin{document}

\begin{center}{\Large \textbf{On the modeling of $t\bar{t}W^\pm$ signatures at the LHC}}\end{center}

\begin{center}
Manfred Kraus
\end{center}

\begin{center}
Physics Department, Florida State University, Tallahassee, FL 32306-4350, USA
\\
* mkraus@hep.fsu.edu
\end{center}

\begin{center}
\today
\end{center}


\definecolor{palegray}{gray}{0.95}
\begin{center}
\colorbox{palegray}{
  \begin{tabular}{rr}
  \begin{minipage}{0.1\textwidth}
    \includegraphics[width=35mm]{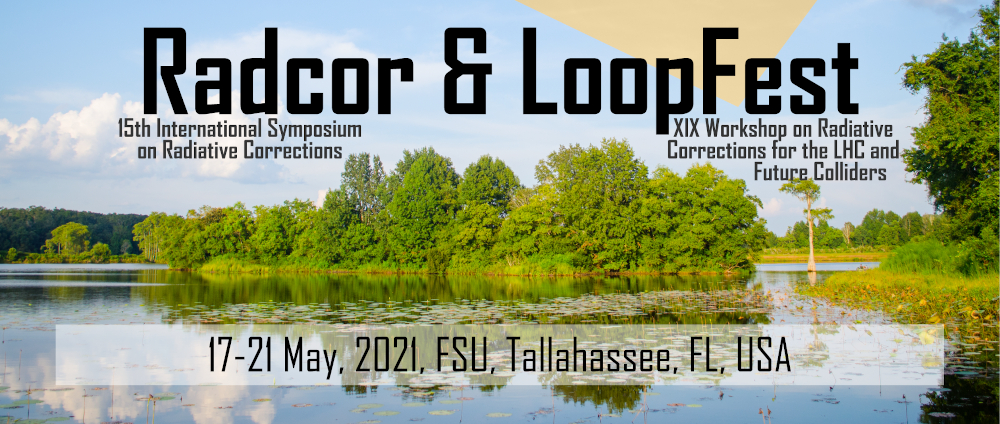}
  \end{minipage}
  &
  \begin{minipage}{0.85\textwidth}
    \begin{center}
    {\it 15th International Symposium on Radiative Corrections: \\Applications of Quantum Field Theory to Phenomenology,}\\
    {\it FSU, Tallahasse, FL, USA, 17-21 May 2021} \\
    \doi{10.21468/SciPostPhysProc.?}\\
    \end{center}
  \end{minipage}
\end{tabular}
}
\end{center}

\section*{Abstract}
{\bf
We discuss our recent progress on improving the theoretical description of the
$pp \to t\bar{t}W^\pm$ process with particular focus on fiducial signatures that
include top-quark decays. We employ a wide range of techniques from parton shower
matching, narrow width approximation to full off-shell computations to assess the
importance of the different modeling approaches.
}

\section{Introduction}
\label{sec:intro}
The production of a $W$ gauge boson in association with a top-quark pair is one
of the rarest and most complex signatures that the Standard Model has to offer.
The multiple resonant decays of top quarks and $W$ bosons yield a plethora of
different final state particles and thus a multitude of different experimental
signatures. In consequence, the $t\bar{t}W^\pm$ process contributes as a dominant
background process to many searches for Beyond the Standard Model (BSM) physics
as well as to the SM measurements of the $t\bar{t}H$ and the $t\bar{t}t\bar{t}$
production process. For the latter processes, the experimental collaborations
have recently reported tensions~\cite{ATLAS:2019nvo} when the $t\bar{t}W^\pm$
process is measured as a background process for multi-lepton signatures.

During the last decade a lot of progress has been made to improve the theoretical
description of the $pp\to t\bar{t}W^\pm$ process. For the inclusive production
with stable top quarks NLO QCD as well as EW corrections are well known and have
been reported in Refs.~\cite{Maltoni:2015ena,Frixione:2015zaa,Frederix:2017wme,
Frederix:2018nkq}. Predictions for fiducial signatures including NLO QCD
corrections to the top-quark decay have been first computed in
Ref.~\cite{Campbell:2012dh} in the Narrow-Width-Approximation (NWA). Only
recently, off-shell effects and non-resonant contributions have been included in
the NLO QCD computation~\cite{Bevilacqua:2020pzy,Bevilacqua:2020srb
,Denner:2020hgg} as well as for EW corrections~\cite{Denner:2021hqi} for
multi-lepton signatures.
Beyond fixed-order, the resummation of soft-gluon emissions has been studied
extensively in Refs.~\cite{Li:2014ula,Broggio:2016zgg,Kulesza:2018tqz,
Broggio:2019ewu,Kulesza:2020nfh} at the inclusive and differential level. The
on-shell $pp\to t\bar{t}W^\pm$ production process has been also matched to parton
showers~\cite{Garzelli:2012bn,Maltoni:2014zpa,Cordero:2021iau} and the impact of
multi-jet merging has been studied for inclusive and fiducial signatures in
Refs.~\cite{Frederix:2020jzp,vonBuddenbrock:2020ter, Frederix:2021agh}.
Furthermore, in Ref.~\cite{Bevilacqua:2021tzp} the approximate combination of
full off-shell effects and parton-shower based predictions has been proposed in
an event generator independent way.

In the following we will summarize our recent efforts to explore the impact of
different approximations made in the description of fiducial signatures for the
$pp\to t\bar{t}W^\pm$ process.

\section{Two same-sign lepton signature}
We start our comparison by considering a two same-sign lepton signature, where we
select events that fulfill
\begin{equation} 
p_T(j) \geq 25~\GeV\;, \qquad |y(j)| < 2.5\;, \qquad 
p_T(\ell) > 15~\GeV\;, \qquad |y(\ell)| < 2.5\;, 
\end{equation}
and have exactly $2$ same-sign leptons, at least $2$ light jets as well as at
least $2b$ jets. The theoretical predictions are obtained using the recent
\powhegbox{} implementation, which is based on one-loop amplitudes provided via
\nlox{}~\cite{Honeywell:2018fcl,Figueroa:2021txg}, \mg5{} and \sherpa{}. We refer
the reader to Ref.~\cite{Cordero:2021iau} for further details of the
computational setup.

\begin{figure}[ht!]
 \centering
 \includegraphics[width=0.49\textwidth]{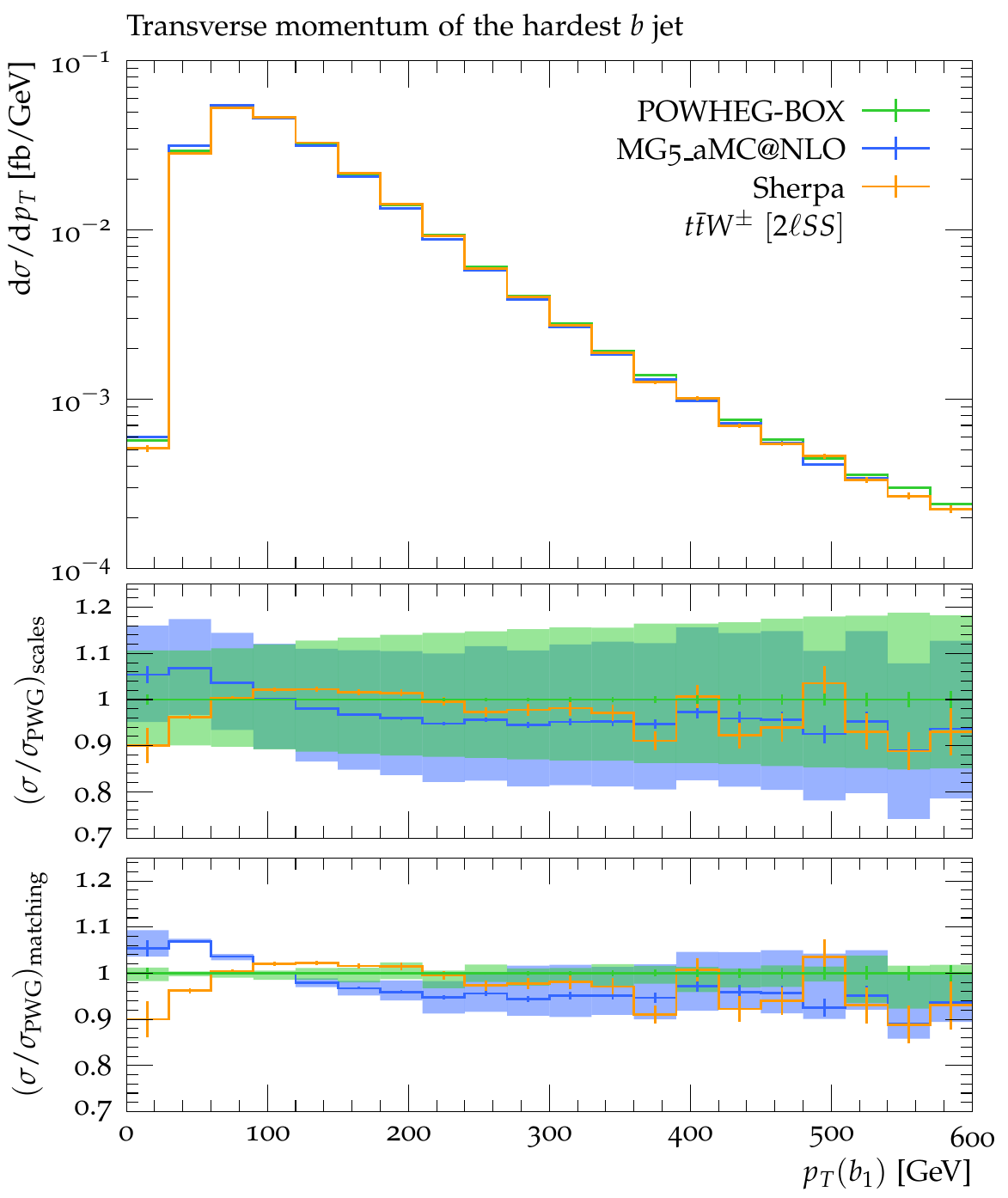}
 \includegraphics[width=0.49\textwidth]{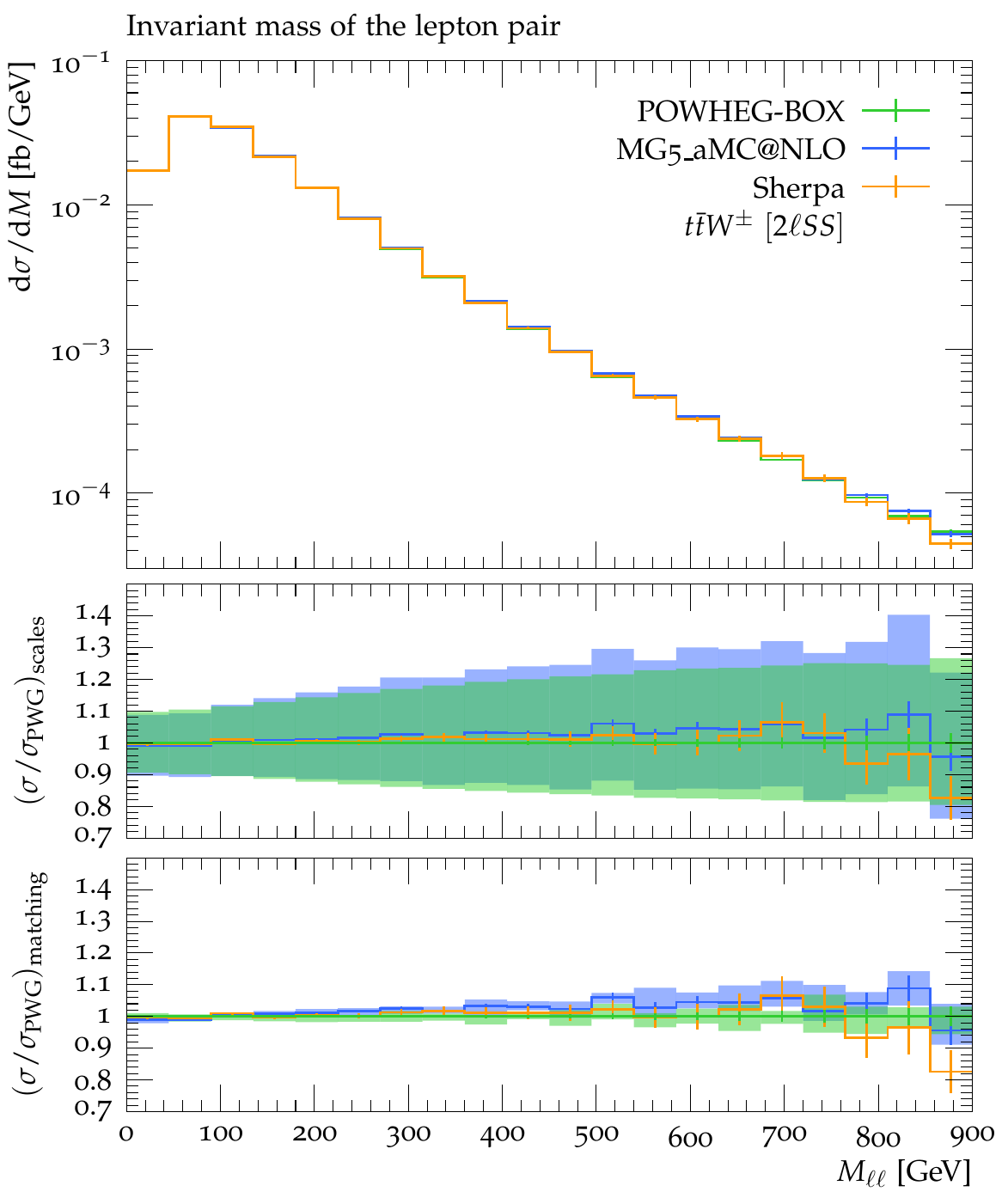}
 \caption{Differential cross section distribution in the two same-sign lepton
fiducial region as a function of the transverse momentum of the hardest $b$ jet
(l.h.s) and of the invariant mass of the lepton pair (r.h.s).}
 \label{fig:PS_1}
\end{figure}
In Fig.~\ref{fig:PS_1} we show the transverse momentum of the hardest $b$ jet and
the invariant mass of the same-sign lepton pair. Both observables are computed
from top-quark decay products, thus allowing to compare the different decay
algorithms employed in the event generators. For the transverse momentum of the
hardest $b$ jet shown on the left we observe good agreement between the various
frameworks with only minor shape modifications below $10\%$. It is well known,
that differences in the treatment of radiation from heavy quarks can account for
these effects. Nonetheless, within the estimated theoretical uncertainties of
$10\%-20\%$ that are dominated by missing higher-order corrections all
predictions are compatible with each other.
In the case of the invariant mass of the two same-sign lepton pair we observe
even better agreement between the considered Monte Carlo event generators. Due to
its leptonic nature the observable is only marginally affected by less than $5\%$
due to the parton shower evolution. Thus, matching uncertainties are negligible
over the whole range. The scale uncertainties start at $10\%$ and increase up to
$25\%$ in the tail of the distribution.

\begin{figure}[ht!]
 \centering
 \includegraphics[width=0.49\textwidth]{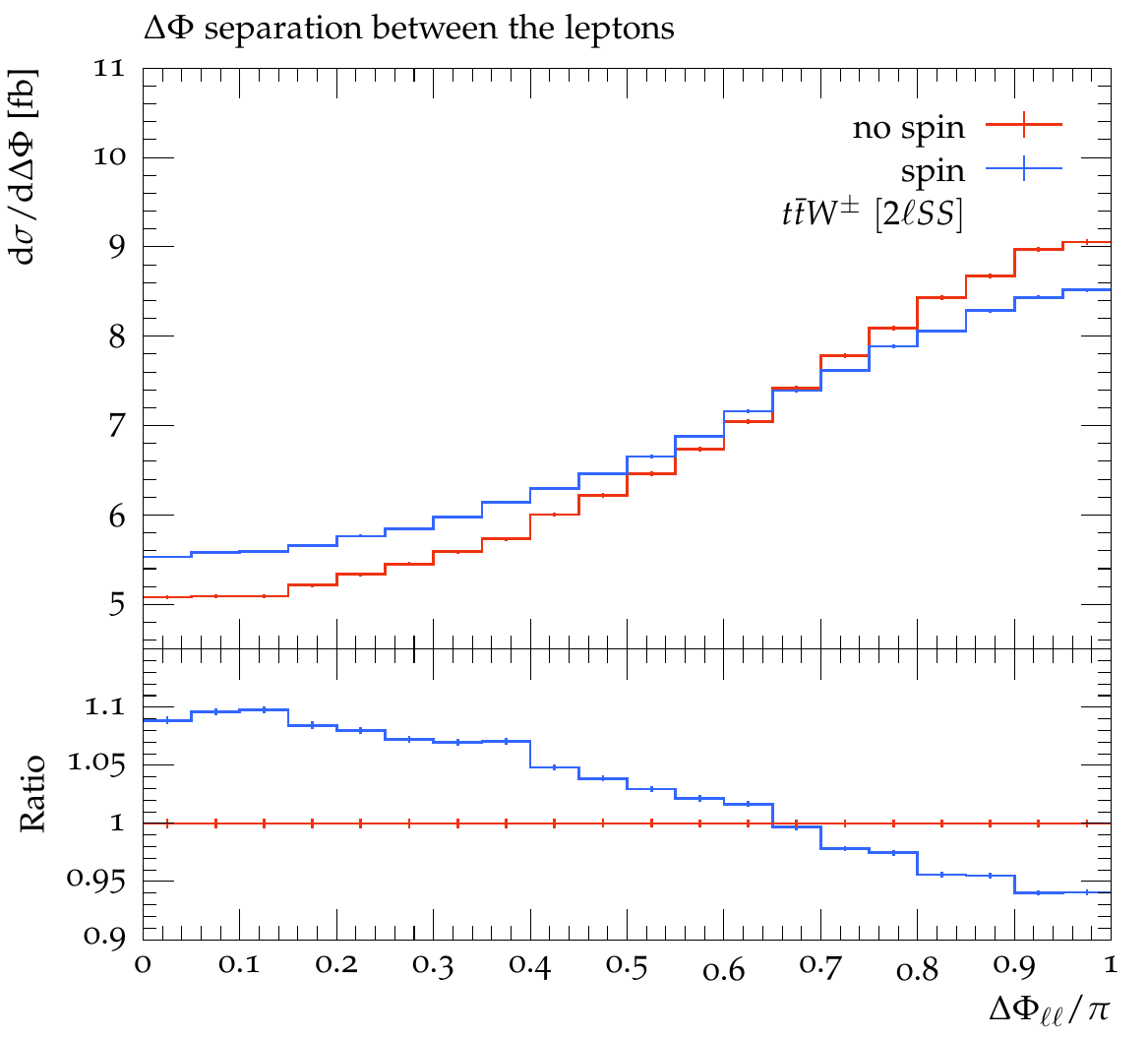}
 \includegraphics[width=0.49\textwidth]{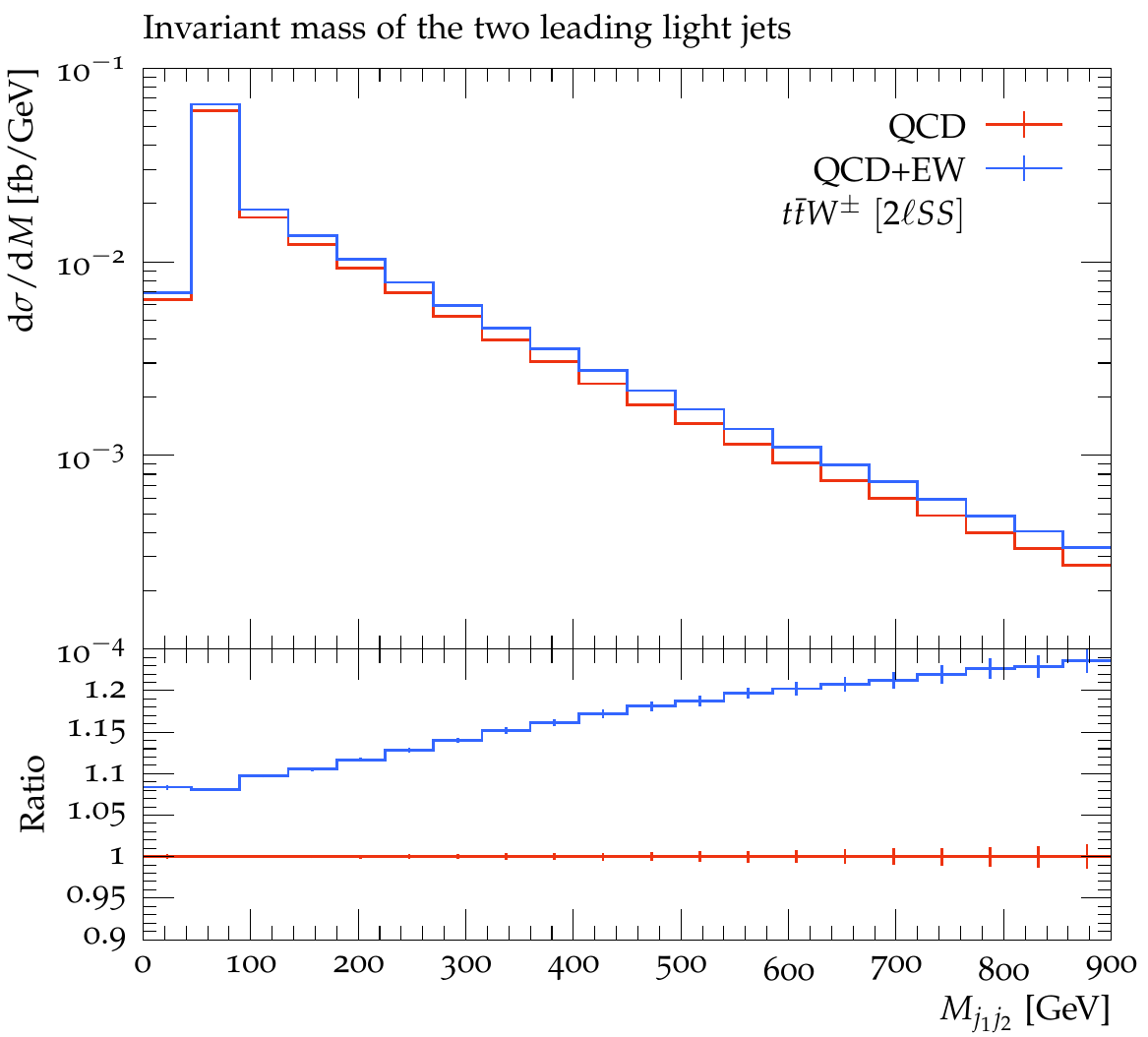}
 \caption{Differential cross section distribution in the two same-sign lepton
fiducial region as a function of the azimuthal angle between the two leptons
(l.h.s) and of the invariant mass of the two hardest light jets (r.h.s).}
 \label{fig:PS_2}
\end{figure}
In Fig.~\ref{fig:PS_2} we show the azimuthal angle between the leptons as well as
the invariant mass of the two hardest light jets. Both observables are sensitive
to the modeling details of the signature. For instance, in the case of the
azimuthal angle between the two same-sign leptons we observe sizable effects due
to spin-correlations in the top-quark decays. The shape of the distribution is
altered at the level of $10\%$. However, for non-leptonic observables we have not
found any indications of spin-correlation effects.
When considering the invariant mass of the two hardest light jets, shown on the
right panel of Fig.~\ref{fig:PS_2}, we observe multiple features. For instance,
the peak of the distribution is centered around the $W$ boson resonance, which
indicates that the majority of events are dominated by jets originating from the
hadronic $W$ decay. This decay, however, is modeled at leading-order accuracy in
all event generators. Including higher-order QCD corrections to these decays is
of utmost importance to achieve a better description of the fiducial signature.
Furthermore, we observe that the impact of the EW contributions increases for 
larger values of the invariant mass because the observable becomes sensitive to
jets generated in the forward region. Nonetheless, for the majority of studied
observables the EW production modes yields a constant $+10\%$ correction at the
differential level.

\section{Multi-lepton signatures}
In the following we compare two different fixed-order approaches to describe
fiducial signatures. We compute the full off-shell process, i.e. the computation
is based on matrix elements for the $pp\to e^+\nu_e\mu^-\nu_\mu e^+\nu_eb\bar{b}$
process.
\begin{figure}[ht!]
 \centering
 \includegraphics[width=\textwidth]{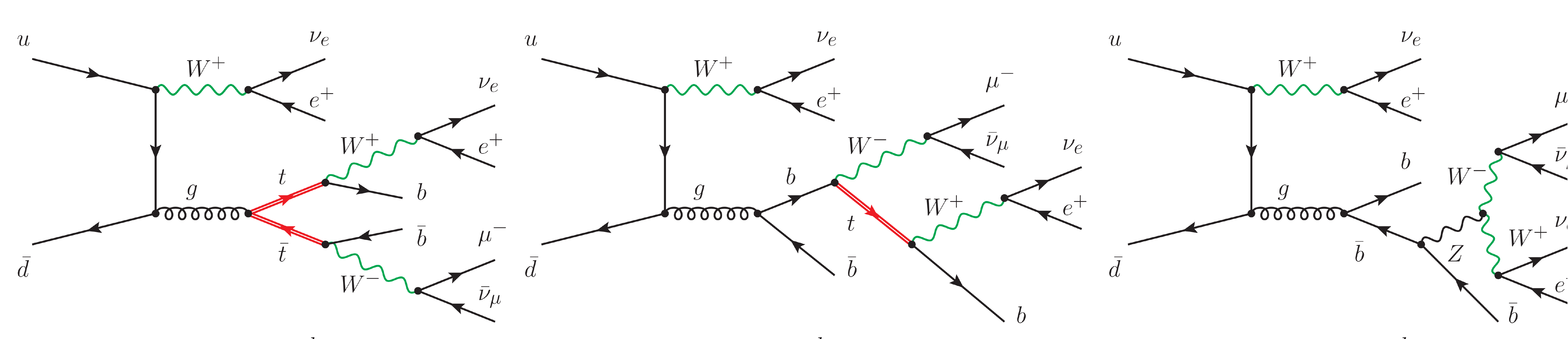}
 \caption{Representative Feynman diagrams for the double, single and non-resonant
contributions to the  $pp\to e^+\nu_e\mu^-\nu_\mu e^+\nu_eb\bar{b}$ amplitude.}
 \label{fig:FD}
\end{figure}
These amplitudes take all double, single and non-resonant contributions together
with all interference effects at the matrix element level into account. In
Fig.~\ref{fig:FD} we show some representative Feynman diagrams for each class of
contribution. The computation is performed in the \helac{}
framework~\cite{Bevilacqua:2011xh} that consists out of
\1loop{}~\cite{vanHameren:2009dr} and
\dipoles{}~\cite{Czakon:2009ss,Bevilacqua:2013iha,Czakon:2015cla} and which has
been already applied to various off-shell processes~\cite{Bevilacqua:2015qha,
Bevilacqua:2018woc,Bevilacqua:2019cvp,Bevilacqua:2021cit}.

On the other hand, the narrow-width-approximation (NWA), which has been recently
automated in our framework~\cite{Bevilacqua:2019quz}, is applied to the process
under consideration to simply greatly the computation. In the NWA only the
leading double-resonant contribution is kept by applying the following relation
to the Breit-Wigner propagators of top quarks and $W$ bosons
\begin{equation}
 \frac{1}{(p^2-m^2)^2 + m^2\Gamma^2} \to \frac{\pi}{m\Gamma}\delta(p^2 - m^2) +
\mathcal{O}\left(\frac{\Gamma}{m}\right)\;.
\end{equation}
Nonetheless, the NWA allows to systematically include NLO QCD corrections in the
production and decay stage with exact spin correlations. For more details on the
computation we refer to Refs.~\cite{Bevilacqua:2020srb,Bevilacqua:2020pzy}.  The
different approaches allow to quantify the importance of NLO QCD corrections to
the decay as well as of the single and non-resonant contributions.

\begin{figure}[ht!]
 \centering
 \includegraphics[width=0.49\textwidth]{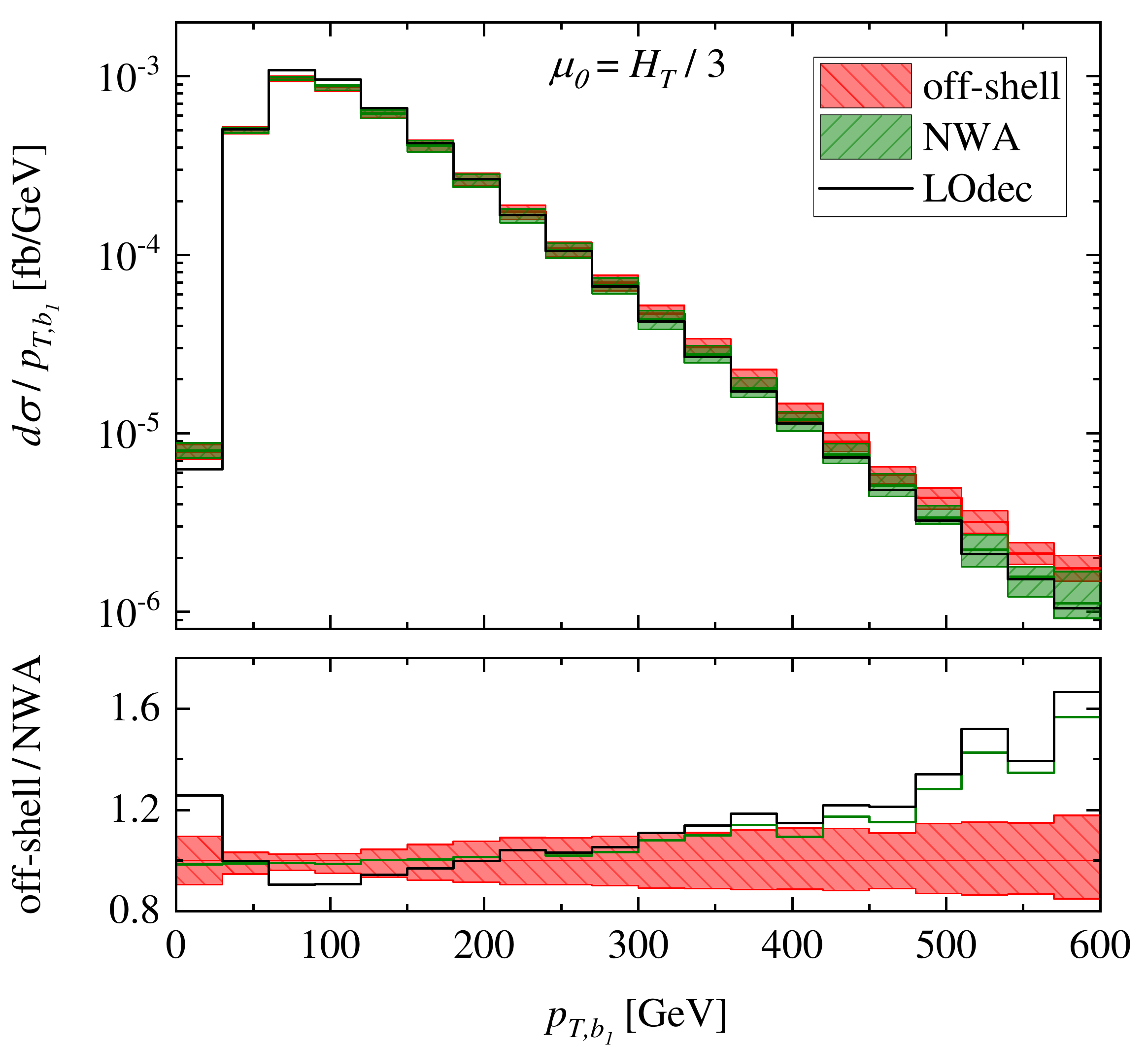}
 \includegraphics[width=0.49\textwidth]{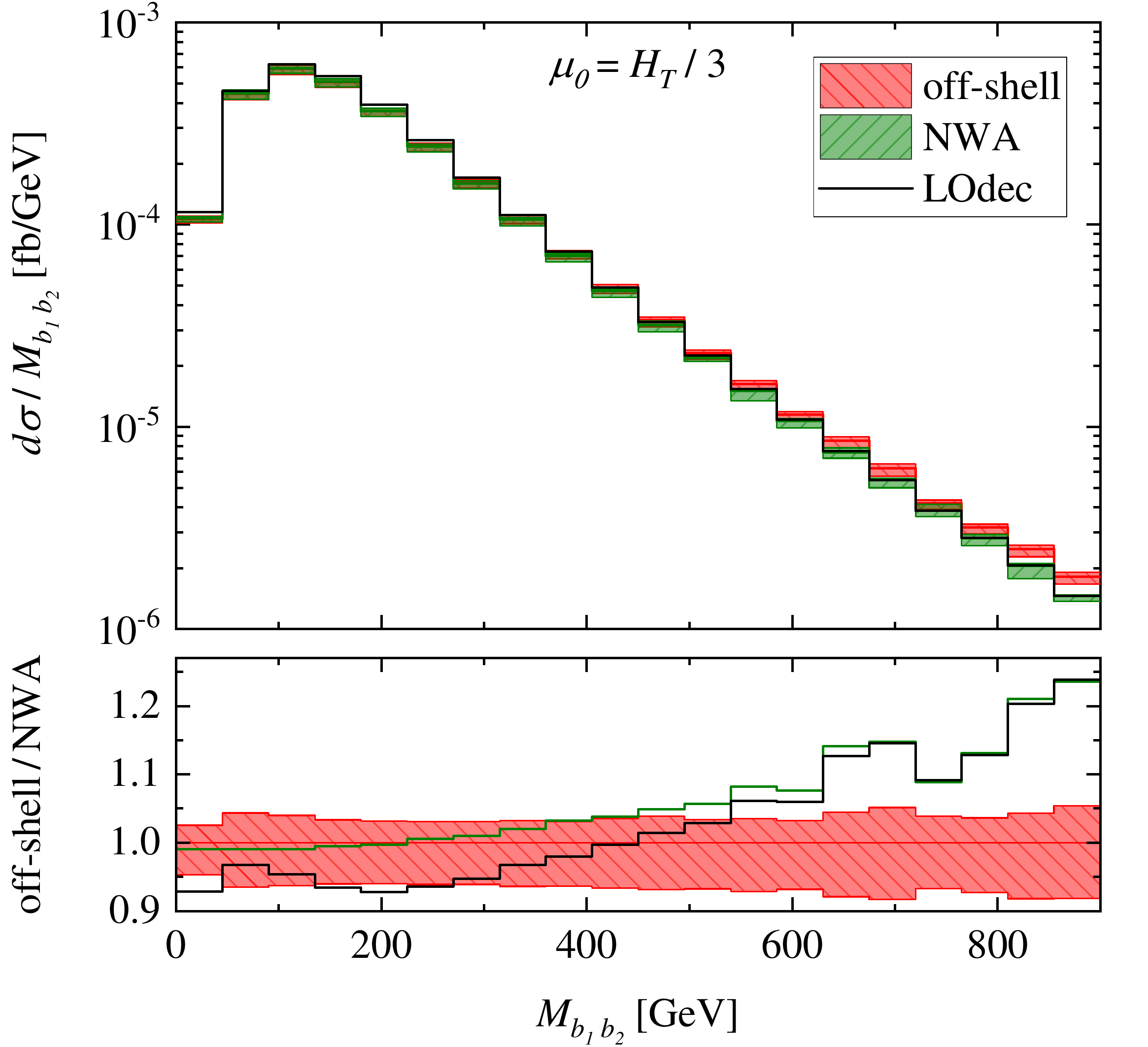}
 \caption{Differential cross section distribution in the three lepton fiducial
region as a function of the transverse momentum of the hardest $b$ jet (l.h.s)
and of the invariant mass of the two hardest $b$ jets (r.h.s).}
 \label{fig:FO_1}
\end{figure}
In Fig.~\ref{fig:FO_1} we illustrate the impact of the various approximations to
the full off-shell computation. For the transverse momentum of the hardest $b$
jet, shown on the left panel, we observe differences up to $60\%$ between the
full off-shell and the full NWA prediction in the tail of the distribution. This
can be attributed to sizable single-resonant contribution in this phase space
region. The size of these effects even outgrow the scale uncertainties that are
below $20\%$. If additionally NLO QCD corrections in the top-quark decays are
omitted in the NWA we observe effects at the $10\%$ level even in the bulk of the
distribution.
Similar effects are observed in the case of the invariant mass of the two hardest
$b$ jets. The size of single-resonant contributions are below $30\%$. However,
also scale uncertainties are smaller and are below $10\%$. The impact of QCD
corrections to the decay is, as in the previous case, at the $10\%$ level.
Therefore, both contributions, i.e. NLO QCD corrections to top-quark decays as
well as single and non-resonant contributions are important for a reliable
description of the fiducial volume.

\begin{figure}[ht!]
 \centering
 \includegraphics[width=0.49\textwidth]{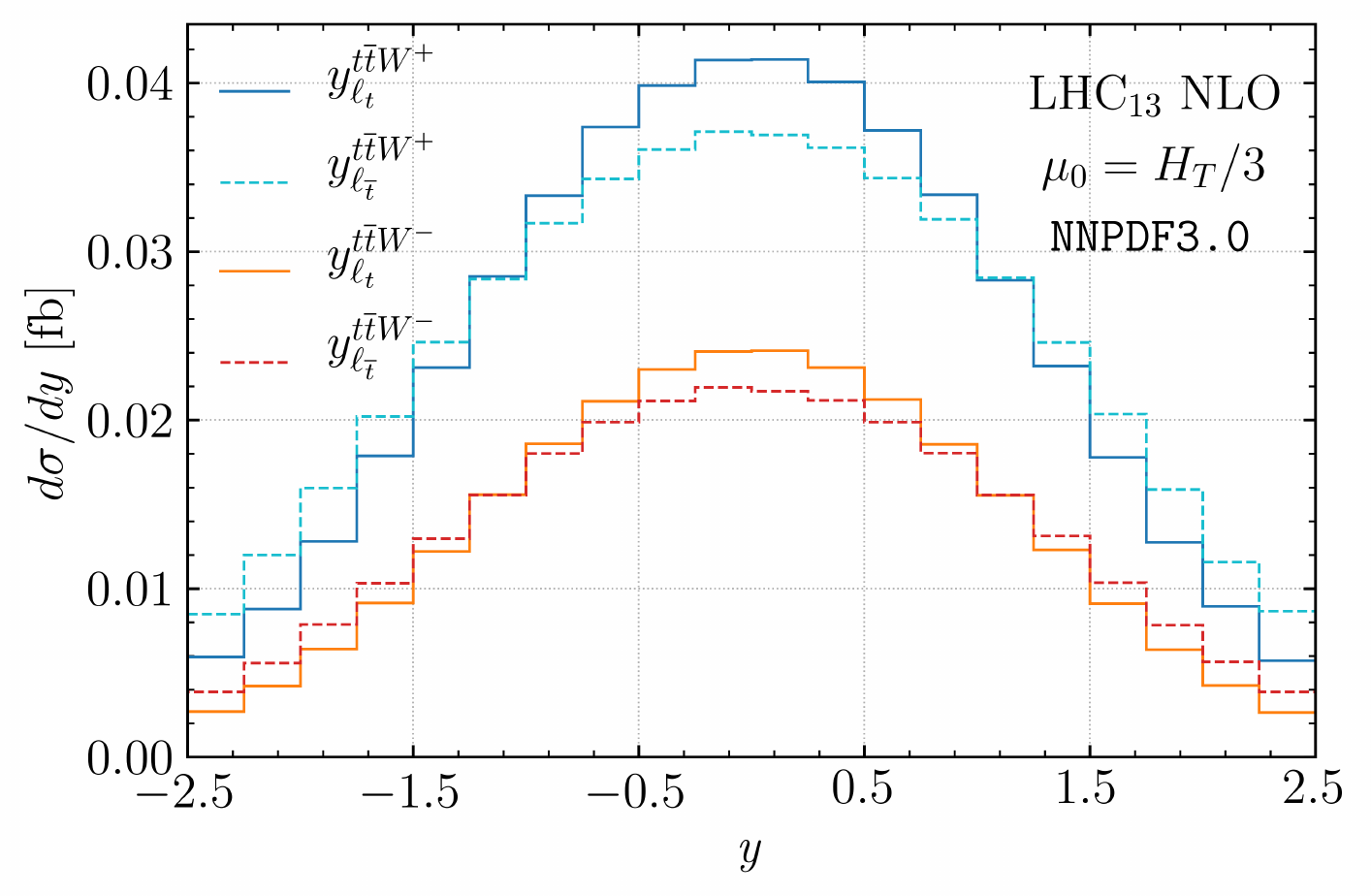}
 \includegraphics[width=0.49\textwidth]{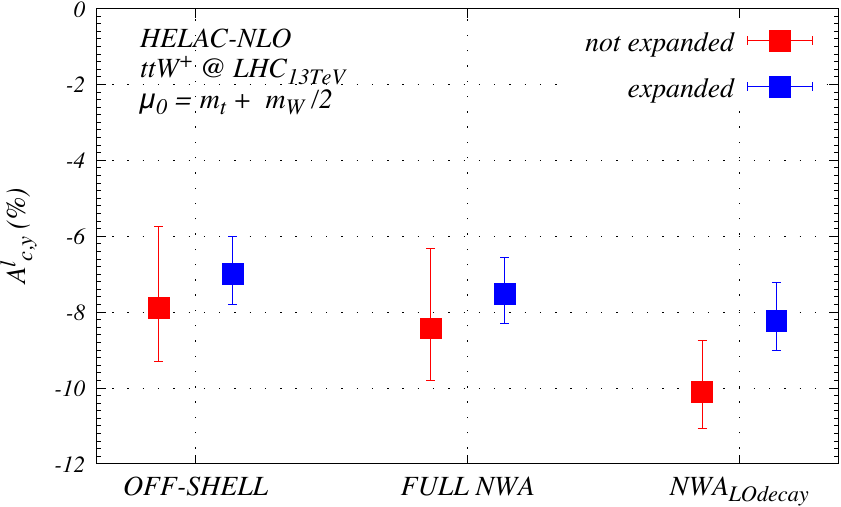}
 \caption{Differential cross section distribution in the three lepton fiducial
region as a function of the rapidity of the lepton originating from the top and
anti-top quark (l.h.s) and the inclusive charge asymmetry for various approaches
to describe the fiducial volume (r.h.s).}
 \label{fig:FO_2}
\end{figure}
We also explored the impact of the various computational approaches in the
context of the charge asymmetry defined via top decay products. On the left panel
of Fig.~\ref{fig:FO_2} we show the rapidity distributions of the lepton
originating from the top and anti-top quark separately for $t\bar{t}W^+$ and
$t\bar{t}W^-$.  It is clearly visible that for each production mode the leptons
are produced asymmetrically. The inclusive charge asymmetry can be computed via
\begin{equation}
 A_c^\ell = \frac{\sigma^+ - \sigma^-}{\sigma^+ + \sigma^-}\;, \qquad 
 \sigma^\pm = \int~d\sigma \theta(\pm\Delta |y|)\;, \qquad \Delta|y| =
|y_{\ell_t}| - |y_{\ell_{\bar{t}}}|\;.
\end{equation}
Our obtained predictions are illustrated on the right panel of
Fig.~\ref{fig:FO_2}. We find that the full off-shell computation predicts a
charge asymmetry of $A_c^\ell = -7.9\%$, while the full NWA gives slightly larger
results of $A_c^\ell = -8.43\%$. If NLO QCD corrections to the decay are omitted
they increase further to $A_c^\ell = - 10.11\%$. Even though all theoretical
predictions are compatible with each other within the estimated uncertainties a
clear trend is visible. In addition, if we expand $A_c^\ell$ to NLO accuracy via
\begin{equation}
 A_{c,exp}^\ell = \frac{\Sigma^-_\lo}{\Sigma^+_\lo}\left( 1 +
\frac{\delta\Sigma^-_\nlo}{\Sigma^+_\lo} -
\frac{\delta\Sigma^+_\nlo}{\Sigma^+_\lo} \right)\;, \qquad \Sigma^\pm = \sigma^+
\pm \sigma^-\;,
\end{equation}
we can estimate the impact of uncontrolled higher-order corrections. Also in this
case we observe that the full off-shell computation shows the smallest dependence
on higher-order terms. We conclude that modeling of the fiducial phase space
volume has a non-negligible effect on the extracted charge asymmetry. In
Ref.~\cite{Bevilacqua:2020srb} we studied more asymmetries also at the
differential level.
\section{Summary \& Outlook}
We studied several aspects of the modeling of fiducial signatures for the
$pp\to t\bar{t}W^\pm$ process at the LHC. In the case of the two same-sign 
lepton signature we compared various parton-shower matched predictions and
explored the impact of subleading EW contributions as well as spin-correlated 
top-quark decays at the differential level. Generally speaking good agreement
between the various event generators have been observed. The subleading EW
production channels contribute at the level of $10\%$. However, their size can
increase up to $25\%$ if the considered observable is sensitive to forward jet
production. Spin-correlation effects have only been found in leptonic
observables, where they can change the shapes of distributions by roughly $10\%$.

On the other hand, for multi-lepton signatures we considered fixed-order
predictions that allows to include also single and non-resonant contributions as
well as NLO QCD corrections to top-quark decays.  By comparing the full off-shell
calculation to the predictions obtained in the NWA we were able to quantify the
impact of those effects at the differential level. We found that corrections in
the top-quark decay have a visible impact at the $10\%$ level even in the bulk of
the distribution. On the contrary, the single and non-resonant contributions can
become sizable in the tail of dimensionful observables and can be as large as
$60\%$.

To improve the theoretical description of fiducial signatures even further a
signature-by-signature approach has to be considered. For multi-lepton signatures
the NNLO QCD corrections to the production part will be of utmost importance.
In the presence of hadronic $W$ decays however the inclusion of NLO QCD
corrections to the decay might be more relevant in the near future.

\subsection*{Acknowledgements}

The author acknowledges support by the U.S. Department of Energy under the grant
DE-SC0010102.
\bibliography{proceedings.bib}

\nolinenumbers

\end{document}